\documentclass[twocolumn,showkeys,showpacs,preprintnumbers,amsmath,amssymb,prb,superscriptaddress]{revtex4}

\usepackage{amsmath}
\usepackage{graphicx}
\usepackage{dcolumn}
\usepackage{bm} 

\newcommand{\DTO}{Dy$_2$Ti$_2$O$_7$}
\newcommand{\HTO}{Ho$_2$Ti$_2$O$_7$}

\begin{document}


\title{\boldmath High-temperature spin relaxation process in 
Dy$_2$Ti$_2$O$_7$ probed by $^{47}$Ti-NQR}

\author{K.~Kitagawa}
\email{kitag@issp.u-tokyo.ac.jp}
\affiliation{Institute for Solid State Physics, University of Tokyo, Kashiwanoha, Kashiwa, Chiba 277-8581, Japan}
\affiliation{Department of Physics, Graduate School of Science, Kyoto 
University, Kyoto 606-8502, Japan}

\author{R.~Higashinaka}
\altaffiliation{Present affiliation: RIKEN (The Institute of Physical and
Chemical Research), Wako, Saitama 351-0198, Japan}
\affiliation{Department of Physics, Graduate School of Science, Kyoto 
University, Kyoto 606-8502, Japan}

\author{K.~Ishida}
\affiliation{Department of Physics, Graduate School of Science, Kyoto 
University, Kyoto 606-8502, Japan}

\author{Y.~Maeno}
\affiliation{Department of Physics, Graduate School of Science, Kyoto University, Kyoto 606-8502, Japan}

\author{M.~Takigawa}
\affiliation{Institute for Solid State Physics, University of Tokyo, Kashiwanoha, Kashiwa, Chiba 277-8581, Japan}

\date{\today}

\begin{abstract}
We have performed nuclear quadrupole resonance (NQR) experiments 
on $^{47}$Ti nuclei in Dy$_2$Ti$_2$O$_7$ in the temperature range 
70 -- 300~K in order to investigate the dynamics of $4f$ electrons with strong
Ising anisotropy.  A significant change of the NQR frequency with temperature
was attributed to the variation of the quadrupole moment 
of Dy $4f$ electrons.  A quantitative account was given by the mean field 
analysis of the quadrupole-quadrupole (Q-Q) interaction in the presence of 
the crystalline-electric-field splitting.   The magnitude and the temperature
dependence of the nuclear spin-lattice relaxation rate was analyzed, 
including both the spin-spin and the Q-Q interactions. The results indicate 
that these two types of interaction contribute almost equally to the fluctuation 
of Dy magnetic moments. 
\end{abstract}

\maketitle

\section{Introduction}
Observation of spin relaxation in magnetic materials provides valuable information 
on the microscopic nature of the interactions between spins. 
When a system of a large spin quantum number has strong Ising anisotropy, the 
mutual spin-flip process is substantially depressed.  The relaxation then becomes 
extremely slow and may eventually lead to spin freezing in real materials with competing
interactions.  A typical example is Dy$_2$Ti$_2$O$_7$, in which Dy$^{3+}$ ions with large 
angular momentum of $J$=15/2 ($4f^{9}$ configuration) form a pyrochlore lattice, 
a network of corner-shared tetrahedra.  

In Dy$_2$Ti$_2$O$_7$ (Ref.~\onlinecite{DTORamirezNature}), 
as well as in other pyrochlore oxides such as Ho$_2$Ti$_2$O$_7$ (Ref.~\onlinecite{HTOHarris}) 
and Ho$_2$Sn$_2$O$_7$\ (Ref.~\onlinecite{HSOMatuhira}), a strong crystalline 
electric field (CEF) forces the magnetic moments of rare earth ions to lie at low 
temperatures along the symmetry axis of CEF, i.e. the [111] or its equivalent directions 
connecting the rare earth site and the center of the tetrahedron it belongs to. 
This strong Ising anisotropy combined with the effective ferromagnetic nearest 
neighbor interaction stabilizes for each bond such magnetic configurations that one moment  
points in and the other points out of the tetrahedron.  Obviously it is not
possible for all bonds to satisfy this condition.  As a compromise, the ground state is selected 
by the rule that two moments point in and other two moments point out of every tetrahedron, 
which is called the ``two-in, two-out'' or the ``ice rule''.  These states are
macroscopically degenerate as long as only the nearest neighbor interaction is considered.  
This state has been named ``spin-ice'' after the $I_h$ phase of 
water ice, which exhibits similar degeneracy with respect to the 
proton configuration.  

The process of spin relaxation in such a system, especially in the spin-ice state, is an
interesting problem.  Although the degeneracy of the spin-ice state may be 
eventually lifted by the long range dipolar interaction, this does not occur in real materials 
within an observable time scale.  Thus the true ground state may never be reached.  
Recently, the relaxation in Dy$_2$Ti$_2$O$_7$ and Ho$_2$Ti$_2$O$_7$ have been 
actively investigated by various experimental means, including the
neutron spin echo experiments in Ho$_2$Ti$_2$O$_7$ (Ref.~\onlinecite{HTOEhlersNSE}), the muon spin rotation measurements in
Dy$_2$Ti$_2$O$_7$ (Ref.~\onlinecite{DTOLagoMuSR}) and the observation of magneto-caloric effects in Dy$_2$Ti$_2$O$_7$
(Ref.~\onlinecite{DTOOrendac}).  However, the time scale of magnetic relaxation 
in Dy$_2$Ti$_2$O$_7$ and Ho$_2$Ti$_2$O$_7$ varies with temperature so widely that no single 
experimental technique can cover the whole range. 

In this paper, we report the results of nuclear magnetic resonance (NMR) and
nuclear quadrupole resonance (NQR) experiments on $^{47}$Ti nuclei in Dy$_2$Ti$_2$O$_7$.
From the analysis of the NMR spectra in high magnetic field and at low temperatures, where the 
configurations of Dy spins are precisely known, we conclude that the magnetic 
hyperfine field at Ti nuclei is entirely due to dipolar interaction with Dy moments.  
The zero-field NQR measurements have been performed in the temperature range 
70 -- 300~K.  Substantial temperature variation of the NQR frequency was 
attributed to the change of the quadrupole moment of Dy $4f$ electrons.  
A quantitative account of the data was obtained by considering both the 
CEF splitting and the quadrupole-quadrupole (Q-Q) interaction in a self-consistent way. 
The fluctuation rate (1/$\tau$) of the Dy magnetic moments extracted from the 
nuclear spin-lattice relaxation rate shows an activated temperature dependence.
The magnitude and the $T$-dependence of 1/$\tau$ was successfully explained
by considering both the spin-spin and the Q-Q interactions, which have comparable  
contributions.  

\section{Experiment}
\begin{figure}[tbp] \centering
\includegraphics[width=0.9\linewidth]{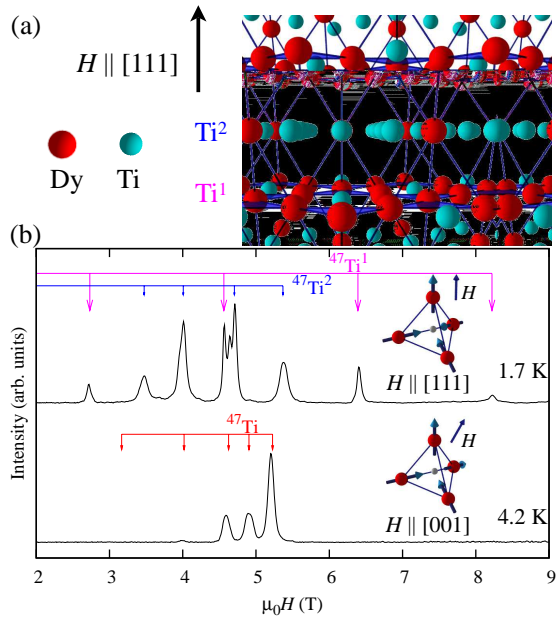} 
\caption{\label{fig:spectra}
(Color online) (a) The crystal structure of Dy$_2$Ti$_2$O$_7$ viewed
perpendicular to the [111] direction.  A magnetic field applied along [111] generates 
two spectroscopically inequivalent Ti sites; Ti$^1$ on the triangular plane and
Ti$^2$ on the Kagom\'e plane of the Ti pyrochlore sublattice. (b)
$^{47}$Ti-NMR field-swept spectra at 9.7~MHz for the two field orientations.  The ``one-in, three-out''
configuration is selected for $H \parallel [111]$, whereas the ``two-in,
two-out'' configuration remains stable for $H \parallel [001]$.
The arrows indicate the calculated resonance positions as described in the text. 
For $H \parallel [001]$, the two resonance lines in low field region were not observed  
probably because of too short $T_{2}$.}
\end{figure}
For the $^{47}$Ti-NQR/NMR experiments we have grown an isotope-enriched single 
crystal of Dy$_2$Ti$_2$O$_7$ using a floating-zone image furnace, starting from the 
sintered polycrystal containing 96\% $^{47}$Ti while the natural abundance of $^{47}$Ti is 7.5\%.
The purpose of this enrichment is to enhance the signal intensity and to avoid overlap 
of resonance lines from $^{47}$Ti and $^{49}$Ti, which have nearly the same 
gyromagnetic ratio ($^{47,49}\gamma/2\pi \simeq 2.400$~MHz/T).  Since the 
demagnetization field of Dy$_2$Ti$_2$O$_7$ is large in high fields and at low temperatures, 
of the order of one tesla, the single crystal was cut into a rod extending
along the $[1{\bar{1}}0]$ direction and the NMR coil was wound only around the
central half of the rod where demagnetization field should be homogeneous.

The standard spin-echo technique was used to obtain the NMR/NQR spectra.  
The nuclear spin-lattice relaxation rate ($1/T_{1}$) was measured by NQR
using the saturation recovery method.   To determine $1/T_{1}$, the intensity of 
the spin-echo signal as a function of the time after the saturation pulse was fit to 
the theoretical formula appropriate for the specific resonance line being
observed\cite{NarathTiNMR}. The spin-spin relaxation rate $1/T_{2}$ was
determined by fitting the spin-echo intensity as a function of the time interval ($\tau$) 
between the $\pi/2$ and the $\pi$ pulses to an exponential function. 

Since the NQR frequencies are relatively low (4 and 9~MHz), the long decay time 
of the electric ringing of the LC resonance circuit after the rf-pulses prevents 
observation of the spin-echo signal for short $\tau$.  This raises a serious problem 
at low temperatures where $T_{2}$ becomes short.  In order to overcome 
this difficulty, we have developed a silicon-diode-based Q-switch.  This device quickly dumps
the quality factor (Q) of the resonance circuit just after the irradiation of rf-pulses, allowing  
us to set $\tau$ as short as to 8~$\mu$s at 9~MHz. 

Still the NQR experiments are not possible in a certain temperature range because of too 
short $T_{2}$ due to the slow dynamics of Ising spins \cite{DTOSnyderNature}.  
The measurements reported in this paper are limited to high temperatures ($\geq 70$~K), 
where the spin dynamics is fast enough, or at low temperatures ($\leq 4$~K) and in high 
fields ($\geq 3$~T), where magnetic moments are completely aligned to certain directions 
by magnetic fields. 

\section{Results and Discussion}
\subsection{High field NMR}
\begin{table}[tb]
\caption{\label{tab:spectra}
Comparisons between the experimental and calculated resonance positions for $^{47}$Ti NMR 
at 9.7~MHz.  The calculated results are obtained by exact diagonalization of 
$\mathcal{H}_\text{nuc}$ [Eq.~\eqref{eq:hnuc}] with or without considering the
hyperfine field from the Dy moments $H_\text{dip}$ [Eq.~\eqref{eq:hdip}].}
\begin{ruledtabular}
\begin{tabular}{lccccc}
& \multicolumn{5}{c}{NMR line positions (T)}\\
\hline
\multicolumn{6}{c}{$H \parallel [111]$, $^{47}$Ti$^1$}\\
Calc. w/o $H_\text{dip}$& 0.382& 2.218& 4.042& 5.872& 7.701\\
Calc. w/ $H_\text{dip}$& 0.138& 2.731& 4.561& 6.391& 8.221\\
Exp.& & 2.72& 4.57& 6.41& 8.22\\
\hline
\multicolumn{6}{c}{$H \parallel [111]$, $^{47}$Ti$^2$}\\
Calc. w/o $H_\text{dip}$& 2.082& 2.693& 3.228& 4.370& 4.761\\
Calc. w/ $H_\text{dip}$& 0.837& 3.473& 4.006& 4.707& 5.358\\
Exp.& & 3.47& 4.00& 4.71& 5.39\\
\hline
\multicolumn{6}{c}{$H \parallel [001]$, $^{47}$Ti}\\
Calc. w/o $H_\text{dip}$& 0.941& 1.536& 3.735& 4.360& 4.423\\
Calc. w/ $H_\text{dip}$& 3.166& 4.018& 4.626& 4.903& 5.225\\
Exp.& & & 4.59& 4.92& 5.21\\
\end{tabular}
\end{ruledtabular}
\end{table}

The crystal structure of Dy$_2$Ti$_2$O$_7$ shown in Fig.~\ref{fig:spectra}(a) contains 
two pyrochlore sublattices, one formed by Ti and the other formed by Dy.  A pyrochlore lattice
can be viewed as the stack of alternating triangular and Kagom\'e planes along [111].  When a 
magnetic field applied is along [111],  The Ti sites on the
triangular planes [Ti$^{1}$ in Fig.~\ref{fig:spectra}(a)] and those on the  Kagom\'e planes
(Ti$^{2}$) become inequivalent, generating distinct NMR lines.
Note that Ti$^1$ (Ti$^2$) is on the Kagom\'e (triangular) planes of the Dy pyrochlore sublattice. 
When the field is applied along [100], all Ti sites are equivalent.   
Figure~\ref{fig:spectra}(b) shows the NMR spectra obtained at the fixed frequency of 9.7~MHz 
for the two field orientations, along $[111]$ and $[100]$, at low enough temperatures.  
The magnetization measurements have established that at these fields and temperatures, 
each Dy moment of 10$\mu_\text{B}$ is completely aligned to a
specific direction\cite{DTOSakakibaraMagnetization, DTOHigashinakaNewPT,
DTOMatsuhiraNewPT,DTOHiroi} as shown in the inset of Fig.~\ref{fig:spectra}(b).
The ``one-in, three-out'' or the ``three-in, one-out'' configuration is selected
for $H \parallel [111]$, while the  ``two-in, two-out'' configuration remains stable for $H \parallel [100]$. 
In the following, we demonstrate that the NMR spectra can be well reproduced by 
considering the dipolar field from Dy moments in these configurations.

The NMR frequencies of $^{47}$Ti are determined by the following 
nuclear hamiltonian, which consists of the Zeeman interaction with the effective magnetic 
field and the nuclear quadrupole interaction with the electric field gradient
(EFG)\cite{SlichterNMR},
\begin{align}\label{eq:hnuc}
\mathcal{H}_\text{nuc} &= -\mu_0\gamma\hbar I_\alpha H^\alpha_\text{eff}
\nonumber\\
&\quad+ \frac{eQ}{6I(2I-1)}V^{\alpha\beta} \left[ \frac{3}{2}
\left(I_\alpha I_\beta + I_\beta I_\alpha\right) - \delta_{\alpha\beta} \bm{I}^2\right].
\end{align}
Here $\bm{I}$ is the nuclear spin with $I$=5/2, $Q\,(=3.02\times
10^{-29}~\text{m}^2)$ is the nuclear quadrupole moment and $V^{\alpha\beta}\,
(=\partial^2V/\partial r_\alpha\partial r_\beta)$ is the EFG tensor, where $V$ is the electrostatic 
potential.  The repeated indices of vectors and tensors should be summed over
the Cartesian coordinates (Einstein's convention).    

The effective field $\bm{H}_\text{eff}$ is the vector sum 
of the external field $\bm{H}$ and the hyperfine field. We assume that the
hyperfine field is given by the dipolar field from Dy moments. 
\begin{align}\label{eq:hdip}
H_\text{dip}^\alpha = \mu_\text{B}g_J
\sum_j &\bigg\{\frac{1}{4\pi}\frac{\partial^2}{\partial r_\alpha\partial
r_\beta}\left(\frac{1}{r}\right) J^{(j)}_\beta \nonumber\\
\quad\quad&- \left(I_\text{D} -
\frac{1}{3}\right)\frac{J^{\alpha;(j)}}{V_A}\bigg\}\quad\text{(SI)},
\end{align}
where $\mu_\text{B}$ is the Bohr magneton, $g_J\,(=4/3)$ is the Lande's $g$
factor, $I_\text{D}$ is the demagnetization factor, and $V_A$ is the atomic
volume per Dy.
The direction of the saturated Dy moments $J$~$(=15/2)$ are fixed
along the local Ising axes in the low-temperature limit of our experiment.
The summation over the spins $j$ should
be cut off within a sphere or performed by the Ewald's method. We take $I_\text{D}$=0.45, which is 
close to the value $1/2$ for a infinitely long cylinder in horizontal fields.
For $\bm{H} \parallel [111]$, the calculated dipolar field is (-0.30, -0.30,
-0.30)~T at Ti$^{1}$ sites and (-0.24, -0.24, -0.66)~T at Ti$^{2}$ sites.  For
$\bm{H} \parallel [001]$ all Ti sites have the same dipolar field (0.18, 0.18,
-0.72)~T.

The EFG tensors can be determined from the NQR measurements at zero magnetic field.
Since the Ti sites have three-fold  rotation symmetry around [111], the EFG is axially 
symmetric, $V_{xx}=V_{yy}=-V_{zz}/2$, where $z \parallel [111]$.  When $\bm{H}_\text{eff} = \bm{0}$,
$\mathcal{H}_\text{nuc}$ can be easily diagonalized to yield two NQR frequencies, 
$\nu_{Q}\,[\equiv 3eQ|V_{zz}|/2I(2I-1)h]$ and 2$\nu_{Q}$, corresponding to the
transitions $I_{z}=\pm 1/2 \leftrightarrow \pm 3/2$ and $I_{z}=\pm 3/2 \leftrightarrow \pm 5/2$, respectively. 

Now we can diagonalize  $\mathcal{H}_\text{nuc}$ numerically to obtain the 
NMR frequencies, which are the difference of the eigenvalues of $\mathcal{H}_\text{nuc}$ 
for various transitions, as a function of external field.  In order to compare the calculated results 
with the experimental NMR spectra obtained by sweeping the field at a constant frequency, this
function should be inverted by Newton's method. The experimental and calculated results for the 
NMR line positions are compared in Fig.~\ref{fig:spectra}(b) and Table~\ref{tab:spectra}.
Here we used the value of $\nu_{Q} = 4.39$~MHz to get the best agreement. This
value is very close to the observed NQR frequency at low temperatures as we
describe below (Fig.~\ref{fig:nuq}).
The results without the hyperfine field are also shown in Table~\ref{tab:spectra}.
The calculated results including the dipolar field agrees extremely well with the experimental data,
demonstrating that the hyperfine coupling in Dy$_2$Ti$_2$O$_7$ is entirely due to the dipolar field.  
The precise knowledge of the hyperfine interaction enables us to make quantitative 
analysis of the NQR relaxation data 
we discuss later.  

\subsection{Zero field NQR}

\begin{figure}[b] \centering
\includegraphics[width=0.9\linewidth]{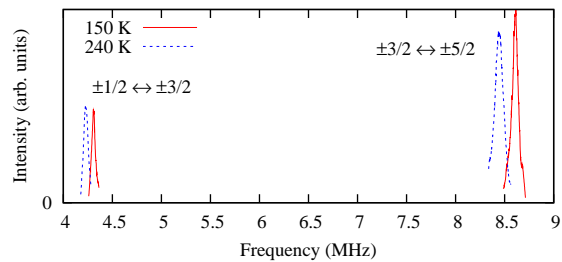} 
\caption{\label{fig:nqrspectra}
(Color online) $^{47}$Ti-NQR spectra at 150~K and 240~K. The NQR frequencies
depend strongly on temperature.}
\end{figure}

\begin{figure}[t] \centering
\includegraphics[width=0.9\linewidth]{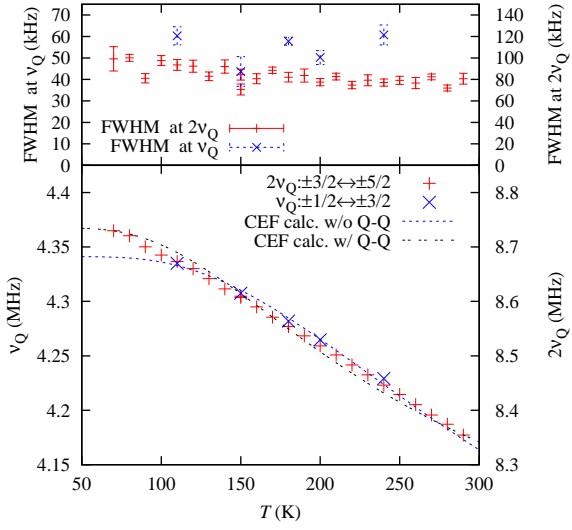} 
\caption{\label{fig:nuq}
(Color online) The peak frequency (lower panel) and the full width at half maximum
(FWHM, upper panel) of the $^{47}$Ti NQR resonance lines for the transitions
$I_{z}=\pm 1/2 \leftrightarrow \pm 3/2$ ($\nu_Q$) and $I_{z}=\pm 3/2 \leftrightarrow \pm 5/2$ (2$\nu_Q$). 
The dotted and double dashed lines in the lower panel show the calculated results considering the 
temperature dependence of the quadrupole moment $\langle C^2_0 \rangle$ without and with the
electric quadrupole-quadrupole interaction, respectively.}
\end{figure}

We now focus on the NQR results in zero field and at high temperatures. 
Figure~\ref{fig:nqrspectra} shows NQR spectra at 240~K and at 150~K. Two resonance lines 
are observed at $\nu_{Q}$ and 2$\nu_\text{Q}$ as expected. However, the value of  $\nu_{Q}$ 
strongly depends on temperatures, which is rather unusual. The temperature
dependence of the peak frequency and the width of the NQR lines are shown in 
Fig~\ref{fig:nuq}. The value of $\nu_\text{Q}$ changes as much as 5\% from 300~K to 70~K. 
This is an order of magnitude larger than what is expected from simple thermal
contraction\cite{DTOLTXRD}. The smooth change of $\nu_\text{Q}$
and the nearly constant line width also rule out any sudden deformation of the crystal structure. 
A natural explanation would be that $\nu_\text{Q}$ 
is influenced by the change of charge density distribution of Dy $4f$ electrons
due to variation of the population among the CEF split $4f$ levels.
For example, Tou~\textit{et\,al.} reported that the temperature dependence of
EFG at Sb nuclei in PrOs$_4$Sb$_{12}$ can be accounted for by the thermal average 
of the hexadecapole moments of Pr$^{2+}$ ($4f^2$) ions which is compatible
with the cubic point symmetry of the Pr sites\cite{SbNQRTou}.
  
\begin{figure}[t]
\centering \includegraphics[width=0.9\linewidth]{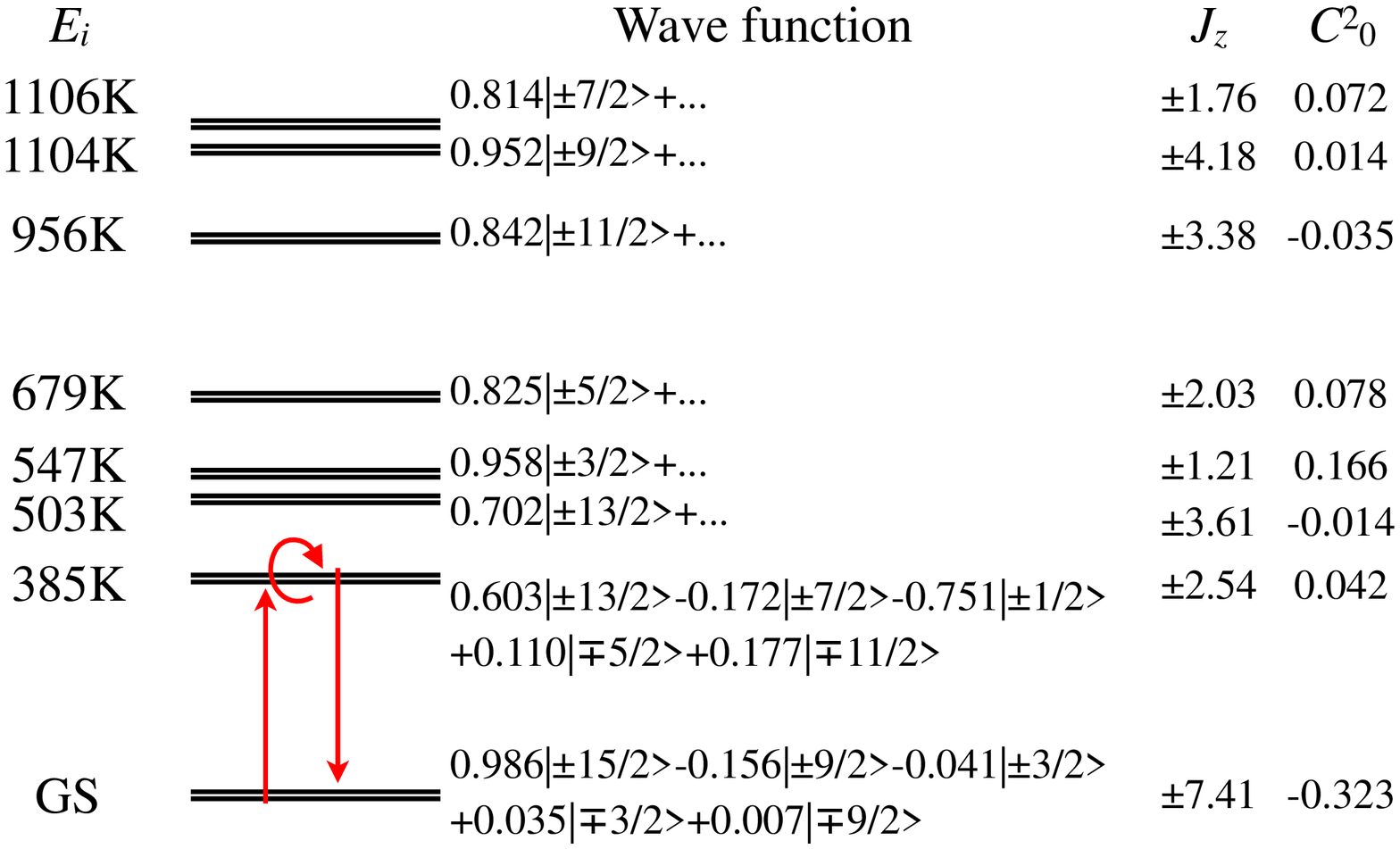} 
\caption{\label{fig:cef}
(Color online) CEF level scheme of \DTO\ inferred from the CEF parameters of
\HTO\ (Ref.~\onlinecite{RosenkranzCEF}). Arrows represent the major process for
the spin-flip relaxations between the ground-state doublet.}
\end{figure}

The CEF Hamiltonian $\mathcal{H_\text{CEF}}$ of single Dy ion in the $D_{3d}$ symmetry
is given by the CEF parameters $B^k_q$ and the normalized
spherical harmonics $C^k_q [=\sqrt{4\pi/(2k+1)}Y^k_q]$ as operators.
\begin{align}
\mathcal{H_\text{CEF}} &=
B^2_0C^2_0 + B^4_0C^4_0 + B^4_3(C^4_3 - C^4_{-3}) \nonumber\\
&+ B^6_0C^6_0 + B^6_3(C^6_3 -
C^6_{-3}) + B^6_6(C^6_6 + C^6_{-6}).
\end{align}
The values of $B^k_q$ for Ho$_2$Ti$_2$O$_7$ have been determined by Rosenkranz and 
co-workers\cite{RosenkranzCEF} by neutron scattering experiments.  
Among the family of lanthanoid titanate pyrochlores, $B^k_q$ should change only modestly.  
We estimate $B^k_q$ in Dy$_2$Ti$_2$O$_7$ by assuming the change of $\langle r^k \rangle$ for the 
$4f$ shells as $\langle r^k \rangle_\text{Dy} / \langle r^k \rangle_\text{Ho}$=
1.05, 1.09, 1.12 (for $k$=2, 4, 6), after Refs.~\onlinecite{FreemanREIons} and \onlinecite{EdvardssonREions}.
In order to calculate the matrix elements of $C^k_q$, the
$|J_z\rangle$ states have to be expanded to the single-electron orbitals with the Clebsch-Gordan
coefficients.
The CEF level scheme for Dy$_2$Ti$_2$O$_7$ obtained by diagonalizing $\mathcal{H_\text{CEF}}$
for the case of $^6H_{15/2}$ multiplet is shown in Fig.~\ref{fig:cef}.
Our results are in good agreement 
with those described in Ref.~\onlinecite{RosenkranzCEF}.  Note that the CEF ground states  
are almost the pure $|J_z = \pm 15/2\rangle$ doublet, leading to the strong Ising anisotropy.  

Since the charge distribution of $4f$ electrons is generally represented by the
electric multipole moments $C^k_q$,  it should be possible to expand $\nu_\text{Q}$ in 
terms of their thermal average $\langle C^k_q \rangle$. Because of the $D_{3d}$ symmetry 
of the Dy ions $\langle C^k_q \rangle$=0 for $q \neq 0$.  Then to the lowest order,
$\nu_\text{Q} = a_{0} + a_{2} \langle C^2_0 \rangle + a_{4} \langle C^4_0 \rangle + 
a_{6} \langle C^6_0 \rangle$.  Although the microscopic mechanism of the coupling 
$a_{k}$ is not understood, it should include the effect of polarization of the surrounding 
medium.  In the following we consider only the quadrupole moment and neglect the 
higher order multipoles,  
\begin{equation}\label{eq:nuq}
\nu_\text{Q} = a_{0} + a_{2} \langle C^2_0 \rangle .
\end{equation}

If we do not consider the quadrupole-quadrupole (Q-Q) interaction between neighboring Dy sites,  
$\langle C^2_0 \rangle$ is simply given by the Boltzman average of the expectation 
values of $C^2_0$ for the CEF eigen states,  
\begin{equation}\label{eq:c20}
\langle C^{2}_0 \rangle = \sum_i \frac{\langle C^{2}_0 \rangle _{i} \exp(-\beta E_i)}{Z},
\end{equation}
where $\beta = (k_\text{B}T)^{-1}$, and $Z = \sum_i \exp(-\beta E_i)$.
The experimental data of $\nu_\text{Q}$ is fit to Eqs.~\eqref{eq:nuq} and \eqref{eq:c20}
with the two adjustable parameters $a_0$ and $a_2$ as shown by the dotted line in Fig.~\ref{fig:nuq} 
(labeled as ``CEF calc. w/o Q-Q'').  Although the experimental data are reproduced 
quite well above 120~K, clear deviation develops at lower temperatures.  Modest change of the 
CEF parameters does not lead to notable improvement.  
In the following we demonstrate that substantial improvement can be achieved
by considering the electric quadrupole-quadrupole (Q-Q) interaction in the mean field
approximation. 

The traceless quadrupole moment $Q_{\alpha\beta}$ in the Cartesian coordinate is
defined as\footnote{The empirical definition of
$Q_{\alpha\beta}$ sometimes appears with a factor of three in literature.},
\begin{widetext}
\begin{equation}
Q_{\alpha\beta} = -e\left(r_\alpha r_\beta  -
\frac{1}{3}\delta_{\alpha\beta} r^2\right) = -\frac{1}{3}e (1-\sigma_2)\langle
r^2
\rangle
  \begin{pmatrix}
  -C^{2}_0 + \frac{\sqrt{6}}{2}(C^{2}_{-2} + C^{2}_2)&
  -\frac{\sqrt{6}}{2}i(C^{2}_{2}-C^{2}_{-2}) & -\frac{\sqrt{6}}{2}(C^{2}_1 -
  C^{2}_{-1})\\ -\frac{\sqrt{6}}{2}i(C^{2}_{2}-C^{2}_{-2}) &
  -C^{2}_0 - \frac{\sqrt{6}}{2}(C^{2}_{-2} + C^{2}_2)&
  \frac{\sqrt{6}}{2}i(C^{2}_1 + C^{2}_{-1})\\ -\frac{\sqrt{6}}{2}(C^{2}_1 -
  C^{2}_{-1})& \frac{\sqrt{6}}{2}i(C^{2}_1 + C^{2}_{-1})& 2C^{2}_0
  \end{pmatrix}. 
\end{equation}
\end{widetext}
In this definition, effects of shielding is already taken into account by the 
Sternheimer shielding factor $\sigma_2 =0.527$ and $\langle r^2\rangle =
0.849$~a.\,u.\cite{EdvardssonREions}. 
From the multipole expansion of the Coulomb potentials\cite{JansenPR}, the
Q-Q interaction between two Dy ions is expressed as
\begin{align}\label{eq:HQQ}
\mathcal{H}_\text{Q-Q}^{12} &=
\frac{1}{2!2!4\pi\epsilon_0}\frac{\partial^4}{\partial r_\alpha\partial
r_\beta\partial r_\gamma\partial r_\delta} \left(\frac{1}{r_{12}}\right)
Q^{(1)}_{\alpha\beta}Q^{(2)}_{\gamma\delta}\nonumber\\ &=\frac{3}{16\pi\epsilon_0}Q^{(1)}_{\alpha\beta}Q^{(2)}_{\gamma\delta}\bigg(
\frac{2}{r^5}\delta^{\alpha\gamma}\delta^{\beta\delta}\nonumber\\
&\quad-\frac{20}{r^7}
r^\alpha r^\gamma\delta^{\beta\delta}
+\frac{35}{r^9}r^\alpha r^\beta
r^\gamma r^\delta \bigg).
\end{align}
It should be noted that in the formulas of the Cartesian coordinates, the operators defined on the
local quantization axes have to be converted as $R Q R^{T}$ and $R \bm J$, where $R$ is the rotation matrix. 

The single site Hamiltonian $\mathcal{H}^{(i)}$ including both CEF and the 
Q-Q interaction in the mean-field approximation can be written as follows.
\begin{align}\label{eq:HMF}
\mathcal{H}^{(i)} &= \mathcal{H_\text{CEF}}^{(i)} + 
\frac{1}{16\pi\epsilon_0} 
\left(C^{2;(i)}_0\langle C^{2}_0 \rangle -
\frac{1}{2}\langle C^{2}_0 \rangle^2\right)\nonumber\\
&\quad\quad\times\left[\sum_{j}\frac{\partial^4}{\partial r_\alpha\partial
r_\beta\partial r_\gamma\partial r_\delta}\left(\frac{1}{r_{ij}}\right)
\frac{\langle Q_{\gamma\delta}^{(j)} \rangle}{\langle
C^{2}_0\rangle}\frac{\langle Q_{\alpha\beta}^{(i)} \rangle}{\langle C^{2}_0
\rangle}\right]\nonumber\\ 
&= \mathcal{H_\text{CEF}}^{(i)} +  (96~\text{K}) \times 
\left(C^{2;(i)}_0\langle C^2_0 \rangle - \frac{1}{2}\langle C^2_0
\rangle^2\right).
\end{align}
The thermal average value of $\langle C^2_0 \rangle$ is calculated again from 
Eq.~\eqref{eq:c20} but now $E_i$ and $\langle C^2_0 \rangle_i$ are the energy 
and the expectation value for the eigenstates of the mean field 
hamiltonian, Eq.~\eqref{eq:HMF}, which itself contains $\langle C^2_0 \rangle$ 
as a parameter.  The temperature dependence of $\langle C^2_0 \rangle$ 
can then be determined self consistently by iteration.  Using this result, 
the experimental data of $\nu_\text{Q}$ is fit to Eq.~\eqref{eq:nuq} as shown 
by the double dashed line in Fig.~\ref{fig:nuq} (labeled as ``CEF calc. w Q-Q'').   
We notice that the calculated result agrees with the experimental data 
much better now, in particular below 100~K.
Although the Q-Q interaction is necessary to consider the
high-temperature excitations, the spin-spin interaction is much more important
in the low-temperature physics since the ground-state (GS) Kramers doublet has
no freedom of multipole moment.

\subsection{Spin dynamics}

\begin{figure}[t]\centering
\includegraphics[width=0.9\linewidth]{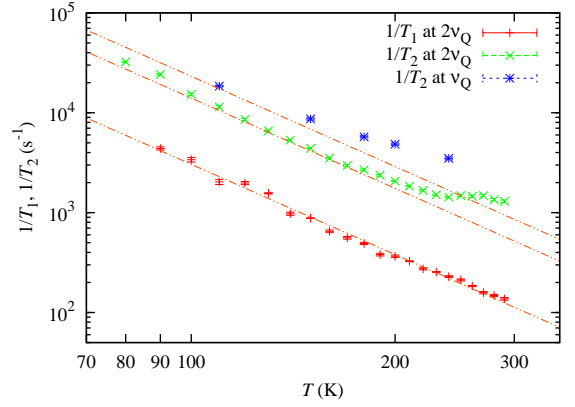} 
\caption{\label{fig:t1t2}
(Color online) Temperature dependences of the spin-lattice relaxation rate $1/T_1$ 
and the spin-spin relaxation rates $1/T_2$ measured for the $I_z = \pm 1/2
\leftrightarrow \pm 3/2$ ($\nu_\text{Q}$) and $I_z = \pm 3/2 \leftrightarrow
\pm 5/2$ (2$\nu_\text{Q}$) transitions. The $1/T_1$ was measured at
2$\nu_\text{Q}$. Three straight lines indicates the $T^{-3}$ dependence. One
is the fit to the data of $1/T_1$ and others are multiplied by 4.6 and 7.6
respectively (see text).}
\end{figure}
Let us now discuss the results of the nuclear spin-lattice relaxation rate $1/T_1$ and the 
nuclear spin-spin relaxation rate $1/T_2$ measured by NQR.  They provide direct
information on the fluctuations of Dy moments at the NQR frequencies.  
Figure~\ref{fig:t1t2} displays the temperature dependence of $1/T_1$ and $1/T_2$.
For the whole temperature range, $1/T_1$ 
obeys the $1/T^3$ behavior remarkably well.
Similar temperature dependence
is observed for $1/T_2$ below 200~K.  For both $1/T_1$ and $1/T_2$, the
relaxation process is considered to be magnetic because relaxation rate
by phononic possesses should increase with increasing
temperature\cite{AbragamNMR}.

The spin-spin relaxation rate is the sum of two contributions, $1/T_2 = (1/T_2)_\text{Dy} + (1/T_2)_\text{nuc}$.
The first term is due to fluctuations of Dy moments and the second term comes from the nuclear spin-spin coupling, 
When the fluctuations of Dy moments are much faster than the NQR frequencies, 
$(1/T_2)_\text{Dy}$ for the transition $I_z=m \leftrightarrow m+1$ is related to $1/T_1$ as
\begin{equation}
\left( \frac{1}{T_2} \right) _\text{Dy} = \frac{1}{T^\parallel_1}\left\{I(I+1) - m(m+1) - 1\right\} +
\frac{1}{T^\perp_1},   
\end{equation}
This relation can be derived by extending the Walstedt's method for the transition 
$I_z=1/2 \leftrightarrow -1/2$ (Ref.~\onlinecite{WalstedtT2}).  Here $T^\parallel_1$ 
($T^\perp_1$) is the relaxation time when the nuclear spin is quantized parallel (perpendicular) 
to $z$, the principal axis of EFG.  Note that $T^\parallel_1$ is equal to
the actual relaxation time measured by NQR experiment.  
The ratio ${T^\parallel_1} / {T^\perp_1}$ is determined by the anisotropy of 
the hyperfine couplings between Dy moments and a Ti nucleus.
\begin{equation}
\frac{T^\parallel_1}{T^\perp_1} = \frac{A_\parallel^2 +A_\perp^2}{2A_\perp^2} =
0.6.
\end{equation}
Here, $A_\parallel = 0.0067$~T$/\mu_\text{B}$ ($A_\perp =
0.019$~T$/\mu_\text{B}$) for the component of the dipolar field from a
single nearest Dy moment parallel (perpendicular) to
$z$. Then, the ratio $(T_2^{-1})_\text{Dy} / T_1^{-1}$ should be 4.6 for 
$m=1/2$ and 7.6 for $m=3/2$.  In fact, the experimental data in
Fig.~\ref{fig:t1t2} agree with these relation below 200~K, indicating that $1/T_2$ is dominated by the 
spin-lattice relaxation process $(T_2^{-1})_\text{Dy}$.  The deviation at higher temperatures 
is likely to be due to the temperature independent term $(1/T_2)_\text{nuc}$.

The rapid increase of $1/T_1$ with decreasing temperature indicates slowing down of 
the fluctuations of Dy moments.  Basically this can be understood from the fact that 
the direct spin flip process is inhibited among the CEF GS doublet,  
which is almost purely the $|J_z = \pm 15/2\rangle$ states.  Thus the spin 
relaxation process at high temperatures should be driven by the transitions 
between the GS and the excited states.  If the effects of the second and higher 
excited states are neglected, the probability of such a process should follow
the Arrhenius-type activated temperature dependence $\exp(-E_\text{a}/k_\text{B}T)$, 
where $E_\text{a}$ is the energy difference between the GS and the first CEF excited states. 
In several experiments to date, the spin relaxation has been fitted to the Arrhenius function: 
the muon spin rotation and the the ac-susceptibility measurements 
($E_\text{a} = 210$~K, Refs.~\onlinecite{DTOLagoMuSR} and
\onlinecite{DTOSnyderNature}), and the nuclear forward scattering experiment
($E_\text{a} = 270$~K, Ref.~\onlinecite{SutterNFS}). Although the $1/T_1$
data in Fig.~\ref{fig:t1t2} apparently does not fit to the Arrhenius function,
 we demonstrate in the following that this can be understood from the
 temperature dependences of both the amplitude of moment fluctuations and the
 fluctuation time $\tau$, and $\tau$ does follow the Arrhenius function. 

\begin{figure}[t]\centering
\includegraphics[width=0.9\linewidth]{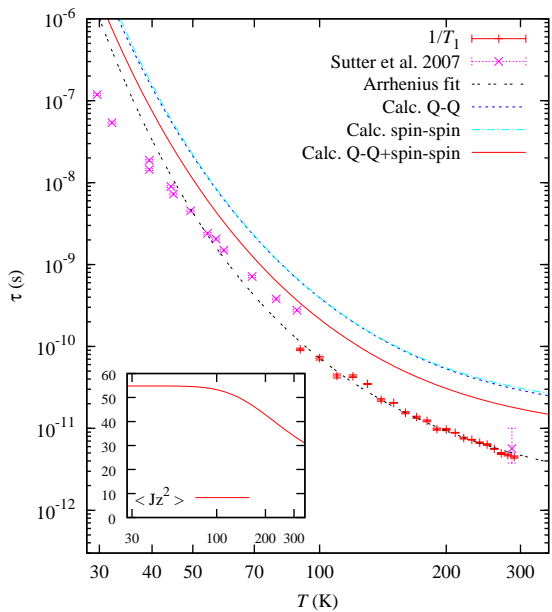} 
\caption{\label{fig:tau}
(Color online) Spin relaxation time $\tau$ as a function of temperature.
extracted from the $^{47}$Ti-NQR relaxation rate $1/T_1$ using Eq.~\eqref{eq:t1tau} 
with the amplitude of the moment fluctuation $\langle J^2 \rangle$ shown in the inset.
The results of the nuclear forward scattering~\cite{SutterNFS} is shown for comparisons. The
Arrhenius fit gives the activation energy $E_\text{a}$ of $409\pm 10$~K. The calculated
results of $\tau$ based on the CEF scheme is shown for three cases, taking account of only the 
spin-spin interaction, only the quadrupole-quadrupole interaction, or both interactions together.   
The results for the first two cases coincide almost completely.}
\end{figure}

When the fluctuations of Dy moments are much faster than the NQR frequencies 
with short enough correlation length, $1/T_1$ is given by the Fermi's golden rule, 
\begin{align}\label{eq:t1tau}
\frac{1}{T_1} &= \frac{z'}{\hbar^2}
\left(^{47}\gamma A_\perp g_J \mu_\text{B}\right)^2 \langle J_z^2 \rangle \tau. 
\end{align}
Here we consider fluctuations of only the $z$ component of Dy moments
because of the strong Ising character of the GS. The temporal correlation of 
$J_z$ is assumed to be described by an exponential function
$\exp(-t/\tau)$ and $z'$ (=6) is the number of nearest neighbor Dy moments.
The amplitude of the moment fluctuations $\langle J_z^2 \rangle$ is calculated
 from the CEF hamiltonian and shown in the inset of Fig.~\ref{fig:tau}.
The fluctuation time $\tau$ can be estimated from Eq.~\eqref{eq:t1tau} by
considering $\langle J_z^2 \rangle$ and the experimental data of $1/T_1$,
as shown in the main panel of Fig.~\ref{fig:tau} for the temperature range 100 -- 300~K.  
In Fig.~\ref{fig:tau}, $\tau$ is found to accommodate to the
simple Arrhenius function with $E_\text{a}$ of $409\pm 10$~K, which is close to
the CEF first excitation level of 385~K.
We also show the temperature dependence of $\tau$ below 100~K obtained by 
Sutter \textit{et\,al.} from the nuclear forward scattering~\cite{SutterNFS}.  
The two sets of experimental results are consistent each other within a factor of three.

The spin flip process relevant to the nuclear relaxation is illustrated by the arrows in
Fig.~\ref{fig:cef}. First, a Dy moment is excited from the GS to one of the the excited doublet.
It then makes the transition to another state within the same excited 
doublet.  This process is caused, for example, by the coupling $J^{(1)}_z J^{(2)}_+$, 
which leaves the spin state at site 1 in the GS unchanged and flips the spin at site 2 in the excited
doublet.  The latter process is allowed for the first excited doublet because it contains the 
$|\pm 1/2 \rangle$ components.  Finally the Dy moment in the excited state relaxes back to the GS 
but final state must be different from from the initial state.  This effectively accomplishes the 
spin flip in the GS: $J_z =15/2 \leftrightarrow -15/2$.  The transition within
the excited doublet is required because otherwise the final state always coincides with the initial  
state.  Below the room temperature, the population of excited states is so small that 
the fluctuation rate is exclusively determined by the first step.

The spin flip rate $1/2\tau$ is then given as the sum of the transition
probability $W$ from the GS to the excited states $i$ when $E_i \gg k_\text{B}T$.
\begin{equation}
\frac{1}{2\tau} \sim \frac{1}{2}\sum_{i(\neq \text{GS})}
W_{\text{GS} \rightarrow i}.
\end{equation}
The factor of $1/2$ is necessary because a half of excited spins return to the
original ground state without flipping.
The transition probability $W_{\text{GS} \rightarrow i}$ is given by the golden rule 
for the mutual transition process. 
\begin{align}\label{eq:W}
W_{\text{GS} \rightarrow i} &\sim z \frac{2\pi}{\hbar}\left|
\left\langle \text{GS}^{(1)}, i^{(2)} \left| \mathcal{H}^{12}
\right| i^{(1)}, \text{GS}^{(2)} \right\rangle \right|^2\nonumber\\
&\quad\quad \times \frac{\exp(-\beta E_i)}{Z}\frac{1}{\pi
\hbar W_{i \rightarrow \text{GS}}},
\end{align}
where the spin at site 2 makes transition from the GS to an excited state $i$ 
while the spin at site 1 does the reverse and  $z (=6)$ denotes the number of 
nearest-neighbor bonds.  We consider the uncertainty principle that the 
energy width of the final state is given by the life time of the excited spin
at site 2, which is $W_{i \rightarrow \text{GS}}$.  From Eq.~\eqref{eq:W} and the
detailed balance relation,
\begin{equation}
W_{\text{GS} \rightarrow i} = W_{i \rightarrow
\text{GS}}\exp(-\beta E_i), 
\end{equation} 
we obtain the expression for $W_{\text{GS} \rightarrow i}$ and $1/\tau$, 
\begin{equation}
W_{\text{GS} \rightarrow i} \sim \frac{\sqrt{2z}}{\hbar}\left| \left\langle
\text{GS}, i \left| \mathcal{H}^{12} \right| i, \text{GS} \right\rangle \right|
\frac{\exp(-\beta E_i)}{\sqrt{Z}},
\end{equation}
\begin{equation}\label{eq:tau}
\frac{1}{\tau} \sim \frac{\sqrt{2z}}{\hbar}\sum_{i(\neq \text{GS})}
\left| \left\langle \text{GS}, i \left|
\mathcal{H}^{12} \right| i, \text{GS} \right\rangle \right|
\frac{\exp(-\beta E_i)}{\sqrt{Z}}.
\end{equation} 
It is noted that Eq.~\eqref{eq:tau} agrees with the Arrhenius behavior if the
the second and the higher CEF levels are ignored.

Finally, we are at the stage to calculate $1/\tau$ in Eq.~\eqref{eq:tau}.  
The interaction between two moments $\mathcal{H}^{12}$, which
is essential to relax the spins, is considered only between the nearest neighbors 
and contains of two terms:
The spin-spin interaction $\mathcal{H}^{12}_\text{spin-spin}$ and the Q-Q interaction
$\mathcal{H}^{12}_\text{Q-Q}$ given in Eq.~\eqref{eq:HQQ}.  The spin-spin
interaction consists of the antiferromagnetic exchange ($J_\text{ex}/3 = -1.24$~K) and
ferromagnetic dipolar interactions according to the Monte Carlo study
of the ``dipolar spin-ice model''~\cite{HertogMonteCarlo,DipolarSpinIce}, 
\begin{align}\label{eq:Hspinspin}
 &\mathcal{H}^{12}_\text{spin-spin}
 \nonumber\\ &= 
 \left\{-\frac{J_\text{ex}}{J^2}\delta^{\alpha\beta} +\frac{\mu_0
 \mu^2_\text{B}g_J^2}{4\pi} \frac{\partial^2}{\partial r_\alpha\partial
r_\beta}\left(\frac{1}{r_{12}}\right)  \right\}J^{(1)}_\alpha J^{(2)}_\beta.
\end{align}

The calculated results of $\tau$ are compared with the experimental results in 
Fig.~\ref{fig:tau}. In the calculated results in, the contributions from the the spin-spin 
interaction and the Q-Q interaction turn out to be nearly the same. When both interactions
are considered, the calculated result of $\tau$ (``Q-Q+spin-spin'' in Fig.~\ref{fig:tau}), 
agrees with the experiment reasonably well.  However, there is still about a factor four difference. 
Possible origin for this may come from effects we have neglected such as the
interactions between second nearest and further neighbors, interactions between
octopoles or higher order multipoles, and short range spin correlation.  However, any of those effects is difficult to estimate
precisely at this moment.

In our formulation thus far, the direct-transition process between the GS is
ignored just because such probability is negligibly small compared with the
process we discussed in the high-temperature region. Nevertheless, the
probability is not strictly zero owing to the existence of $\pm 9/2$ and
$\pm 3/2$ components in the GS (Fig.~\ref{fig:cef}). We consider that the direct
spin flip process has to be responsible for the excitations in the spin-ice state of \DTO\ below 1~K, which is still not uncovered experimentally.

\section{Conclusion}
We have investigated the Dy Ising-spin relaxation process in a range of
70 -- 300~K using $^{47}$Ti-NMR/NQR in Dy$_2$Ti$_2$O$_7$. The hyperfine
coupling, CEF level scheme, and interaction Hamiltonians,
were deduced from the low-temperature NMR spectra, experimental CEF of
 Ho$_2$Ti$_2$O$_7$ (Ref.~\onlinecite{RosenkranzCEF}), and the temperature dependence of the NQR
frequency, respectively. Those information enabled us
to calculate the absolute values of the Dy spin-flip rate and NQR relaxation rate.
The absolute value of the
NQR relaxation rate was analyzed in the CEF level scheme with two types of
mutual spin-flip Hamiltonians: the spin-spin and Q-Q interaction. The Q-Q
interaction is also necessary to account for the feature of the modification in
the NQR frequency. Our experimental and calculated results reveals the spin-flip
process and the importance of the Q-Q interaction in the high-temperature
nature of the Ising-spins in Dy$_2$Ti$_2$O$_7$.

\begin{acknowledgments}
This work was in
part supported by the Grants-in-Aid for Scientific Research from JSPS and MEXT
of Japan and by the 21COE program ``Center for Diversity and
Universality in Physics'' from MEXT of Japan.
\end{acknowledgments}


\bibliography{resubmit} 

\end{document}